\documentclass[a4paper,fleqn,usenatbib]{mnras}

\usepackage{newtxtext}

\usepackage[T1]{fontenc}
\usepackage{ae,aecompl}

\usepackage{graphicx}
\usepackage{amsmath}	
\usepackage{amssymb}	

\usepackage{color}
\title[Evidence of an evolved nature of MWC\,349A]{Evidence of an evolved nature 
of MWC\,349A\thanks{Based on observations obtained at the Gemini Observatory, which is operated by the Association of Universities for Research in Astronomy, Inc., under a cooperative agreement with the NSF on behalf of the Gemini partnership: the National Science Foundation (United States), National Research Council (Canada), CONICYT (Chile), Ministerio de Ciencia, Tecnolog\'{i}a e Innovaci\'{o}n Productiva (Argentina), Minist\'{e}rio da Ci\^{e}ncia, Tecnologia e Inova\c{c}\~{a}o (Brazil), and Korea Astronomy and Space Science Institute (Republic of Korea) under program IDs 
GN-2013A-Q-78 and GN-2013B-Q-11.}}

\author[M. Kraus et al.]{M.~Kraus$^{1}$\thanks{E-mail: michaela.kraus@asu.cas.cz},
        M.~L.~Arias$^{2,3}$\thanks{Member of the Carrera del Investigador Cient\'{i}fico, CONICET}, L.~S.~Cidale$^{2,3}$\footnotemark[3], A.~F.~Torres$^{2,3}$\footnotemark[3]
            \\
\\
$^{1}$ Astronomical Institute, Czech Academy of Sciences, 
Fri\v{c}ova 298, 251\,65 Ond\v{r}ejov, Czech Republic\\
$^{2}$ Departamento de Espectroscop\'ia Estelar, Facultad de Ciencias
Astron\'omicas y Geof\'isicas, Universidad Nacional de La Plata \\ 
$^{3}$ Instituto 
de Astrof\'isica de La Plata, CCT La Plata, CONICET-UNLP, Paseo del
Bosque s/n, B1900FWA, La Plata, Argentina
}

\date{Accepted XXX. Received YYY; in original form ZZZ}

\pubyear{2019}

\begin{document}
\label{firstpage}
\pagerange{\pageref{firstpage}--\pageref{lastpage}}
\maketitle

\begin{abstract}
The Galactic emission-line object MWC\,349A is one of the brightest radio stars 
in the sky. The central object is embedded in an almost edge-on oriented 
Keplerian rotating thick disc that seems to drive a rotating bipolar wind. The 
dense disc is also the site of hot molecular emission such as the CO bands 
with its prominent band heads in the near-infrared spectral range. 
Despite numerous studies, the nature of MWC\,349A is still controversial with 
classifications ranging from a pre-main sequence object to an evolved 
supergiant.  We collected new high-resolution near-infrared spectra in the $K$- 
and $L$-bands using the GNIRS spectrograph at Gemini-North to study the 
molecular disc of MWC\,349A, and in particular to search for other molecular 
species such as SiO and the isotope $^{13}$CO. The amount of $^{13}$CO, obtained 
from the $^{12}$CO/$^{13}$CO ratio, is recognized as an excellent tool to 
discriminate between pre-main sequence and evolved massive stars.
We find no signatures of SiO band emission, but detect CO band emission with 
considerably lower intensity and CO gas temperature compared to previous 
observations. Moreover, from detailed modelling of the emission spectrum, 
we derive an isotope ratio of $^{12}$CO/$^{13}$CO = 4$\pm$1. Based on
this significant enrichment  of the circumstellar environment in $^{13}$CO we 
conclude that MWC\,349A belongs to the group of B[e] supergiants, and we discuss
possible reasons for the drop in CO intensity.
\end{abstract}


\section{Introduction}

The emission-line object MWC\,349A ($\alpha$ = 20h\,32m\,45.53s, $\delta$ = 40\degr 39\arcmin 36\farcs 60) is one of the brightest radio 
stars in the sky \citep*{1972Natur.240..230B}. Radio continuum observations 
suggest that MWC\,349A is driving an ionised wind with an outflow velocity of 
$25-50$\,km\,s$^{-1}$ \citep*{1981A&A....93...48A} and a mass-loss rate of 
$10^{-5}$\,M$_{\sun}$yr$^{-1}$ \citep{1975A&A....39..217O}. The lack of 
photospheric absorption lines hampers significantly the assignment of a proper 
classification, but the numerous \ion{He}{i} lines in emission speak in favour 
of a hot central source with a spectral type between late O 
\citep*{1980ApJ...239..905H} and about B0 \citep{2002A&A...395..891H}. Estimates
for the object's distance and luminosity range from 1.2\,kpc 
\citep{1985ApJ...292..249C} to 1.7\,kpc \citep*{2002AJ....123.1639M}, and from 
$3\times 10^{4}$\,L$_{\sun}$ \citep{1985ApJ...292..249C} to  $8\times 
10^{5}$\,L$_{\sun}$ \citep{2012A&A...541A...7G}, respectively. 
This distance range is consistent with the parallax and its 
error provided by Gaia DR2 \citep{2018A&A...616A...1G}.
The central object suffers from an extinction of $A_V \simeq 10$\,mag, of which 
about two magnitudes seem to be circumstellar in origin 
\citep{1985ApJ...292..249C}. 

The morphology of the radio emission at 6\,cm is spherical 
\citep{1985ApJ...292..249C}, whereas at 2\,cm it reveals a bipolar shape of the 
ionised wind with an east-west extent smaller than 0\farcs 4, indicating an 
equatorial disc seen edge-on \citep{1985ApJ...297..677W}. 
The star has a strong infrared (IR) excess emission \citep{1970ApJ...161L.105G}
indicating large amounts of circumstellar dust, and a dusty disc oriented in 
east-west direction was resolved by infrared interferometric 
\citep*{2001ApJ...562..440D} as well as by speckle interferometric observations 
\citep{1983A&A...120..237M, 1986A&A...155L...6L}. IR images at 24\,$\mu$m taken 
with the {\sl Spitzer Space Telescope} show an extended (over several 
arcminutes) infrared double-cone (or X-shaped nebula) structure 
\citep{2012A&A...541A...7G, 2013ApJ...777...89S}. Moreover, many optical and IR 
emission lines of elements in diverse ionisation states display double-peaked 
profiles \citep*{1996A&AS..118..495A, 2016MNRAS.456.1424A}. These have been 
suggested to form within a Keplerian rotating neutral gas disc with an ionised 
surface layer \citep[e.g.,][]{1986ApJ...311..909H}. 

Additional support for a rotating circumstellar disc is provided by the numerous 
hydrogen recombination maser and laser lines for which MWC\,349A is famous. The 
first maser lines at mm wavelengths have been discovered by 
\citet{1989A&A...215L..13M}, and since then, many more maser and laser 
transitions have been identified that spread from the far-infrared up to the 
millimetre range \citep{1994A&A...283..582T, 1994A&A...288L..25T, 
1998A&A...333L..63T, 1996Sci...272.1459S}. Modelling of the maser lines reveals 
that their profiles consist of two components: one in agreement with Keplerian 
rotation within a narrow circumstellar ring, and another one suggesting a wind 
emanating from and co-rotating with the disc \citep[e.g.][]{2011A&A...530L..15M, 
2013A&A...553A..45B, 2014A&A...571L...4B, 2017ASPC..508..279B, 
2017ApJ...837...53Z}.

MWC\,349 has been proposed to be a binary system, in which the companion, a 
B0 III star (MWC\,349B), is located about $2\farcs 4$ west of the massive and 
luminous B[e] component MWC\,349A \citep{1985ApJ...297..677W}. However, recent 
analysis of the radial velocities of the two objects seems to speak against a 
gravitationally bound system, which reopens the discussion whether MWC\,349A 
belongs to the Cygnus OB2 association  \citep{2017ApJ...851..136D}.
 
Despite numerous studies about MWC\,349A, the nature of this enigmatic object 
remains elusive. Classifications range from a massive pre-main sequence object 
due to the disc-bipolar wind shape seen on small scales on radio images and 
traced by the maser and laser lines \citep[e.g.,][]{1985ApJ...292..249C, 
2013ApJ...777...89S} to an evolved massive star such as a B[e] supergiant  
\citep{1980ApJ...239..905H, 2002A&A...395..891H} or a luminous blue variable 
\citep{2012A&A...541A...7G} based on the huge, up to 5\,pc scale infrared nebula 
structure associated with the star. To solve the issue of the unclear 
evolutionary state of MWC\,349A, once and for all, a trustworthy age indicator 
is needed. 

As was shown by \citet{2009A&A...494..253K}, the most ideal tool to discriminate 
between a young, pre-main sequence and an evolved nature of a massive star is 
provided by the abundance ratio between $^{12}$C and its isotope $^{13}$C 
measured in terms of the molecular abundance ratio $^{12}$CO/$^{13}$CO within 
their circumstellar environments. In the pre-main sequence stage, the 
$^{12}$C/$^{13}$C ratio has the interstellar value of $\sim 90$ 
\cite[see, e.g.,][]{2012A&A...537A.146E}. During the evolution of a massive 
star, the surface abundance ratio changes considerably, and the value can drop 
to $^{12}$C/$^{13}$C $<$ 10 or even $^{12}$C/$^{13}$C $<$ 5, depending on the 
initial mass of the star and its rotation which mixes chemically processed 
material to the surface. Once on the surface, the isotopes are transported via 
mass-loss in form of stellar winds and mass ejections to the environment. There, 
they can be bound in $^{12}$CO and $^{13}$CO, in case the physical parameters in 
the environment, in terms of density and temperature, favour the condensation of 
molecules. 

Circumstellar discs provide the most ideal environments in which molecules can 
form and survive in substantial amounts. Within the discs, the molecules are 
shielded from the dissociating stellar ultraviolet radiation field. Emission 
from hot CO gas has been detected in the IR spectra of both accretion discs of 
pre-main sequence stars \citep[e.g.,][]{1987ApJ...312..297G, 
1989ApJ...345..522C, 1995Ap&SS.224...25C, 2010MNRAS.408.1840W, 
2013MNRAS.429.2960I, 2018MNRAS.477.3360I} and discs formed from the outflows and 
ejecta of evolved massive stars such as the B[e] supergiants 
\citep{1988ApJ...324.1071M, 1988ApJ...334..639M, 1989A&A...223..237M, 
1996ApJ...470..597M, 2012A&A...543A..77W, 2016A&A...593A.112K, 
2017AJ....154..186K, 2018A&A...612A.113T, 2018MNRAS.480..320M, 
2018MNRAS.480.3706K}. Enrichment in $^{13}$CO has been found, based on 
measurements of the $^{12}$CO/$^{13}$CO ratio in high-quality near-IR spectra, 
solely in the discs of confirmed evolved objects \citep{2010MNRAS.408L...6L, 
2013A&A...558A..17O, 2013A&A...549A..28K, 2014ApJ...780L..10K}, reinforcing the 
power of this isotope-ratio method for the discrimination between young and 
evolved stars \citep{2015AJ....149...13M}.

CO band emission from MWC\,349A was discovered by \citet{1987ApJ...312..297G}. 
Although their spectral range covered the positions of the first $^{13}$CO band 
head, the identification of the isotope was uncertain due to the very low 
resolution ($R\sim 650$) of their spectra. Follow-up observations in the 
$K$-band had either similarly low resolution, or were limited to studies of the 
first band heads of $^{12}$CO to determine the kinematics within the molecular 
emission region \citep{2000A&A...362..158K}. 

In this work, we wish to solve the long-standing issue of the unclear nature of 
MWC\,349A. For an unambiguous classification as either a pre-main sequence 
object or an evolved massive star, we acquired new high-quality near-IR spectra, 
which allow us to measure precisely the $^{12}$CO/$^{13}$CO ratio in its 
circumstellar disc. Moreover, we aim to search for emission from 
SiO. The detection of emission from hot SiO gas would allow us to derive 
complementary information about the physical properties within the molecular 
disc of MWC\,349A \citep[see][]{2015ApJ...800L..20K}.

\begin{figure*}
\begin{center}
\includegraphics[width=\hsize,angle=0]{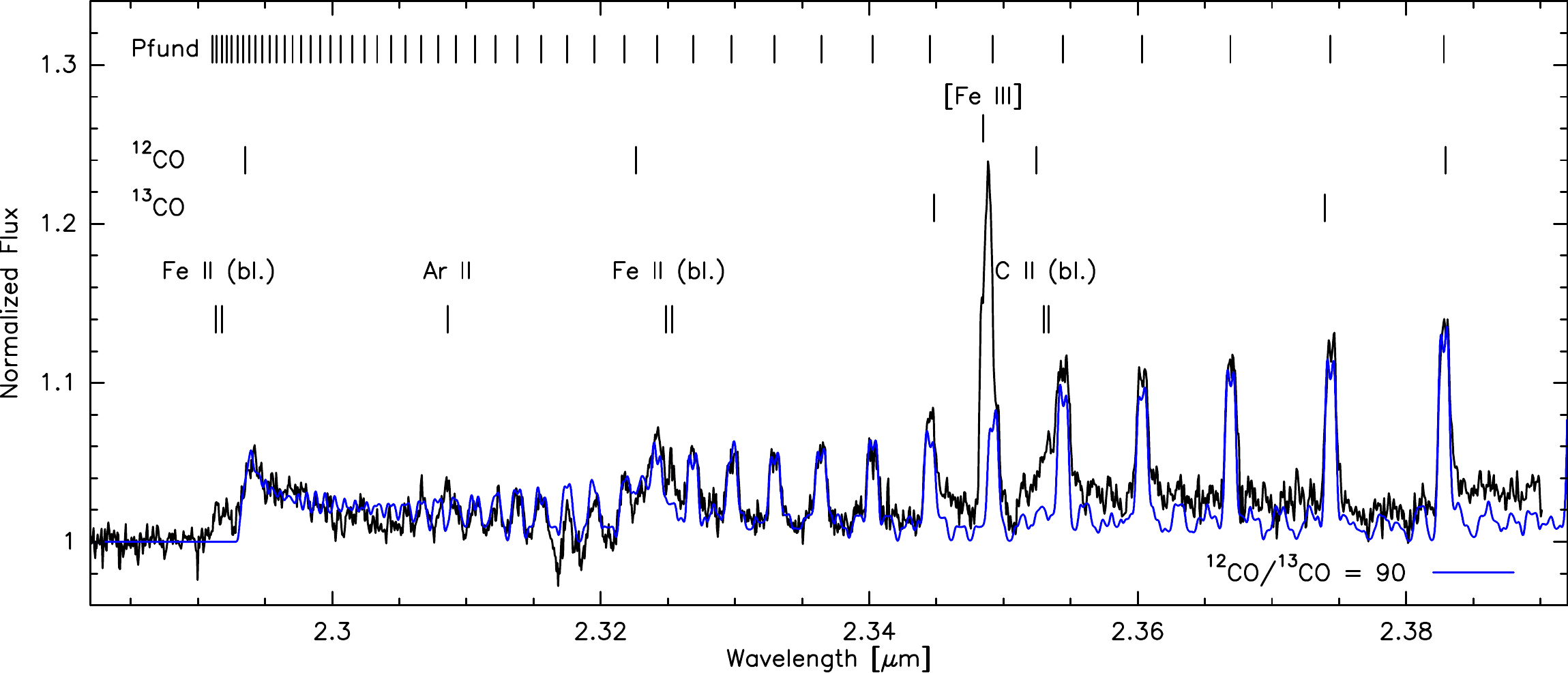}
\caption{Normalized high-resolution $K$-band spectrum (black) of MWC\,349A, 
showing the most prominent emission lines and the positions of the band heads of 
$^{12}$CO and $^{13}$CO. The best-fitting model (blue) is also shown. It is the 
superposition of the emission from the Pfund series  and the CO-band emission
for an isotope abundance ratio of 90 (interstellar value). See 
Sect.\,\ref{sect:results} for details.}
\label{fig:K-band-obs}
\end{center}
\end{figure*}

\begin{figure}
\begin{center}
\includegraphics[width=\hsize,angle=0]{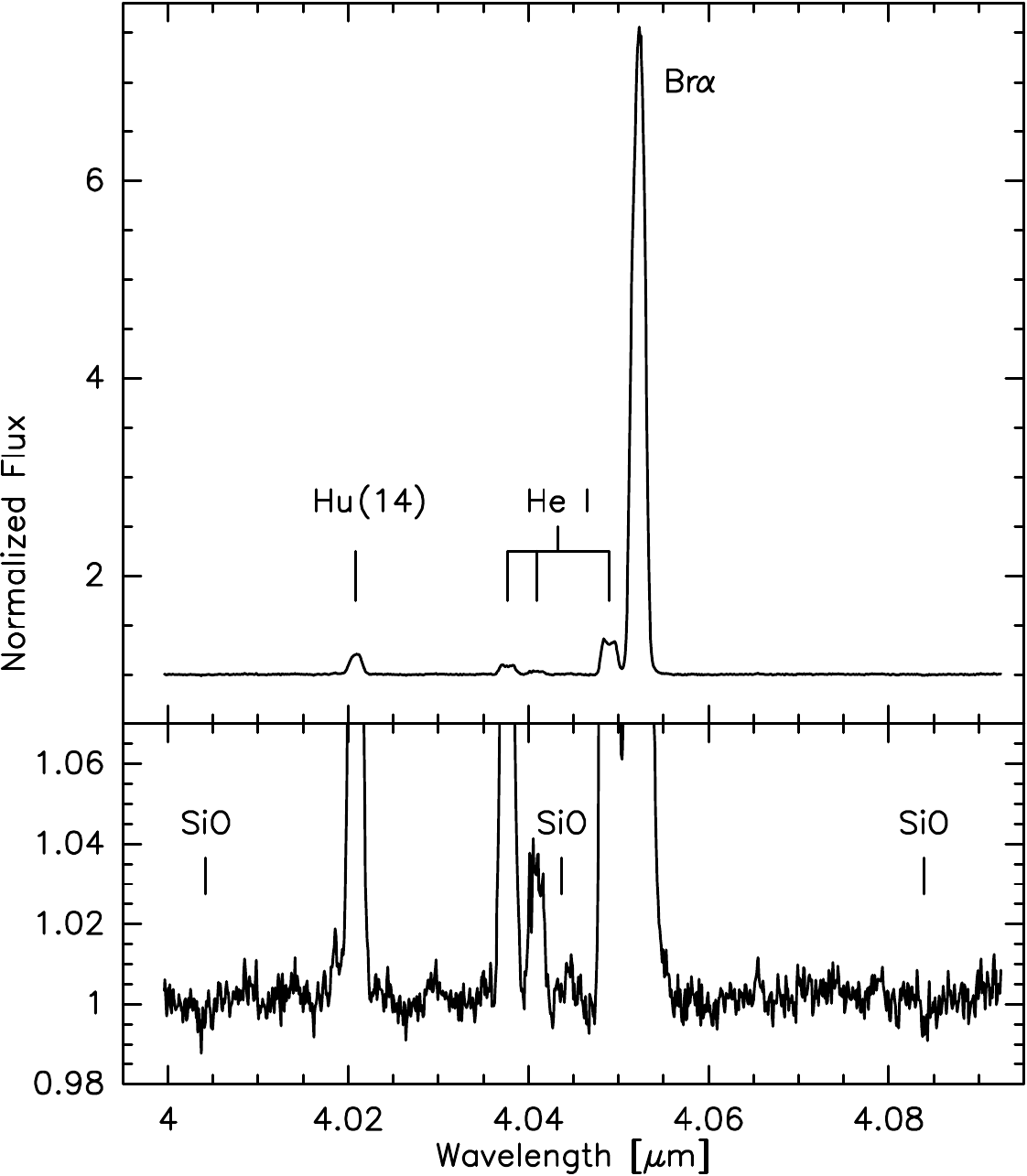}
\caption{High-resolution $L$-band spectrum of MWC\,349A with prominent emission lines labeled (top). The bottom panel is a zoom around the continuum, indicating the wavelengths of the SiO band heads.}
\label{fig:L-band-obs}
\end{center}
\end{figure}

\section{Observations and reduction}

A medium-resolution ($R \sim$ 6\,000) $K$-band spectrum was acquired on 2013 
July 7, and high-resolution ($R \sim$ 18\,000) $K$- and $L$-band spectra on 
2013 August 18. The observations were carried out using the Gemini Near-InfraRed 
Spectrograph \citep[GNIRS,][]{2006SPIE.6269E..4CE} at GEMINI-North under Program 
IDs GN-2013A-Q-78 and GN-2013B-Q-11. For each science acquisition, a telluric standard star 
was observed close in time and airmass.


The medium-resolution spectrum was taken with the 110.5\,l\,mm$^{-1}$ grating, 
the short camera, and the $0\farcs 3$ slit. It covers the wavelength range from 
2.280 to 2.475\,$\mu$m. 

The high-resolution spectra were taken with the 110.5\,l\,mm$^{-1}$ grating, the 
long camera ($0\farcs 05$\,pix$^{-1}$) and the $0\farcs 1$ slit. In the $K$-band 
the spectra were centred on 2.312\,$\mu$m and 2.36\,$\mu$m to trace the 
molecular emission of the CO isotopes $^{12}$CO and $^{13}$CO, and in the 
$L$-band on 4.05\,$\mu$m to cover the region of the SiO bands.

Exposures for the target and the telluric standard star were taken in two 
positions along the slit (A and B) with a series of ABBA sequences.
Data reduction was performed with standard 
IRAF\footnote{IRAF is distributed by the National Optical Astronomy Observatory, 
which is operated by the Association of Universities for Research in 
Astronomy (AURA) under cooperative agreement with the National Science 
Foundation.} tasks. The procedure includes subtraction of AB pairs, 
flat-fielding, and telluric correction. The wavelength calibration was 
performed using the telluric lines. The high-resolution $K$- and 
$L$-band spectra achieved signal-to-noise ratios of $\sim$130 and $\sim$280, 
respectively. The final spectra were normalized to the continuum. 

The medium and high-resolution $K$-band spectra, taken with a separation of 
about one month, show no noticeable difference in the appearance of the 
emission features. Therefore, the following investigation and analysis is based 
purely on the high-resolution spectrum.

\section{Results}\label{sect:results}

\begin{table}
\centering
\caption{Identification of emission lines besides the lines Pf\,(24) to Pf\,(62) 
of the Pfund series.}
\label{tab:line-ids}
\begin{tabular}{ccccc}
\hline 
\multicolumn{2}{c}{$K$-band} & & \multicolumn{2}{c}{$L$-band} \\
$\lambda$ & Element & & $\lambda$ & Element  \\
($\mu$m) & & & ($\mu$m)  &\\
\hline 
  2.2913  &   Fe\,{\sc ii}    &  & 4.0209   &     Hu\,(14) \\
  2.2918  &   Fe\,{\sc ii}    &  & 4.0377   &     He\,{\sc i} \\
  2.2935  &   $^{12}$CO (2-0)   &  & 4.0409   &     He\,{\sc i} \\
  2.3086  &   Ar\,{\sc ii}    &  & 4.0490   &     He\,{\sc i} \\
  2.3227  &   $^{12}$CO (3-1)   &  & 4.0522   &     Br\,$\alpha$ \\
  2.3248  &   Fe\,{\sc ii}    &  &          & \\   
  2.3253  &   Fe\,{\sc ii}    &  &          & \\ 
  2.3448  &   $^{13}$CO (2-0)   &  &          & \\  
  2.3485  &   [Fe\,{\sc iii}] &  &          & \\ 
  2.3525  &   $^{12}$CO (4-2)   &  &          & \\ 
  2.3530  &   C\,{\sc ii}     &  &          & \\
  2.3533  &   C\,{\sc ii}     &  &          & \\
  2.3739  &   $^{13}$CO (3-1)   &  &          & \\
\hline 
\end{tabular}
\end{table}

Figures\,\ref{fig:K-band-obs} and \ref{fig:L-band-obs} show the observed 
high-resolution spectra of MWC\,349A in the $K$- and $L$-band, respectively. The 
$K$-band displays emission lines from the Pfund series ranging from 
Pf\,(24) up to Pf\,(62), weak CO band emission, as well as emission from a few 
metallic lines. Their wavelengths and identifications are given in columns 1 and 
2 of Table\,\ref{tab:line-ids}, respectively. The line 
identification is based on wavelengths retrieved from the NIST Atomic Spectra 
Database\footnote{https://www.nist.gov/pml/atomic-spectra-database}, except for 
the line [Fe\,{\sc iii}]\,$\lambda$2.3485 that was taken from 
\citet{1996MNRAS.281..493L}. The wavelengths of the CO band heads were taken 
from \citet{2000A&A...362..158K}.

The only lines we could identify in the $L$-band are from hydrogen and 
He\,{\sc i} (see columns 3 and 4 of Table\,\ref{tab:line-ids}). No traces of 
emission from the SiO molecule are found (see bottom panel of 
Fig.\,\ref{fig:L-band-obs}). First-overtone band emission of SiO 
has been detected so far from the high-density environments of four B[e] stars
\citep{2015ApJ...800L..20K}, and in all four cases, the intensity in the SiO 
bands was 5--15 times lower than the intensity in the first-overtone CO bands  
published by \citet{2018MNRAS.480..320M}. Considering that such a trend might 
also apply to MWC\,349A, the weakness of its CO band emission with only 
$\sim$5--6\% of the continuum flux might be the reason for the absence of 
detectable SiO band emission.

\subsection{Pfund line emission}

The high-resolution $K$-band spectrum (Fig.\,\ref{fig:K-band-obs}) shows that 
the Pfund lines display slightly double-peaked profiles, implying rotational 
motion of the ionised gas. To model the Pfund line emission, we utilize the code 
developed by \citet{2000A&A...362..158K} that computes the emission of the 
series according to Menzel case B recombination, assuming that the lines are 
optically thin.  

For the computations of the line intensities, we fix the electron temperature at 
10\,000\,K, which is a suitable value for the ionised gas around a hot OB-type 
star. Fixing the temperature has no significant impact on the shape of the lines 
or the relative line intensities of the individual Pfund lines. The observed 
$K$-band spectrum implies a cut-off in the Pfund series indicating pressure 
ionisation. The maximum number of the series, Pf(62), delivers a hydrogen 
density of $(3.5\pm 0.3) \times 10^{12}$\,cm$^{-3}$ within the line-forming 
region. Considering a partially ionised gas, this value poses an upper limit for 
the electron density. We use this value for the computation of the Pfund line 
spectrum. For such a high density, the emission is in local thermodynamic 
equilibrium (LTE).

For the profile function, we assume that the emission of the Pfund lines emerges 
from a narrow ring of gas with constant temperature and density revolving the 
star. This assumption is justified based on the double-peaked profiles that we 
observe for the Pfund lines. The inclination angle of the system is fixed at 
80$\degr$ under the convention that an edge-on orientation corresponds to an 
inclination angle of 90$\degr$. This choice of 80$\degr$ has been made to be in 
line with the model considerations for the CO band emission (see below). Then, 
the only free parameters for the computation of the Pfund line emission spectrum 
are the rotation velocity, $v_{\rm rot, Pf}$, and a Gaussian component 
$v_{\rm gauss, Pf}$. This Gaussian component combines the broadening 
contributions from a possible turbulent motion of the gas (or a possible wind 
component) and the contribution of the thermal motion. The final Pfund emission 
spectrum is convolved with the resolution of the spectrograph.

We obtain reasonable fits to the observed line profiles for a rotation velocity 
of 45$\pm$5\,km\,s$^{-1}$ in combination with a Gaussian component of 
20$\pm$5\,km\,s$^{-1}$. This high value for the Gaussian velocity is needed to 
reproduce the line wings and the shallowness of the double peak structure. 
Considering a contribution of 12--13\,km\,s$^{-1}$ for the thermal 
motion we derive a turbulent component of about 15$\pm$5\,km\,s$^{-1}$.

\subsection{CO band emission}

Our spectrum spreads over a wide wavelength range, covering the positions of the 
first four band heads of $^{12}$CO and the first two band heads of $^{13}$CO 
(Fig.\,\ref{fig:K-band-obs}). However, due to the strong contamination with the 
Pfund emission lines, only the first two band heads of $^{12}$CO can be easily 
identified by eye in the spectrum.

\begin{figure}
\begin{center}
\includegraphics[width=\hsize,angle=0]{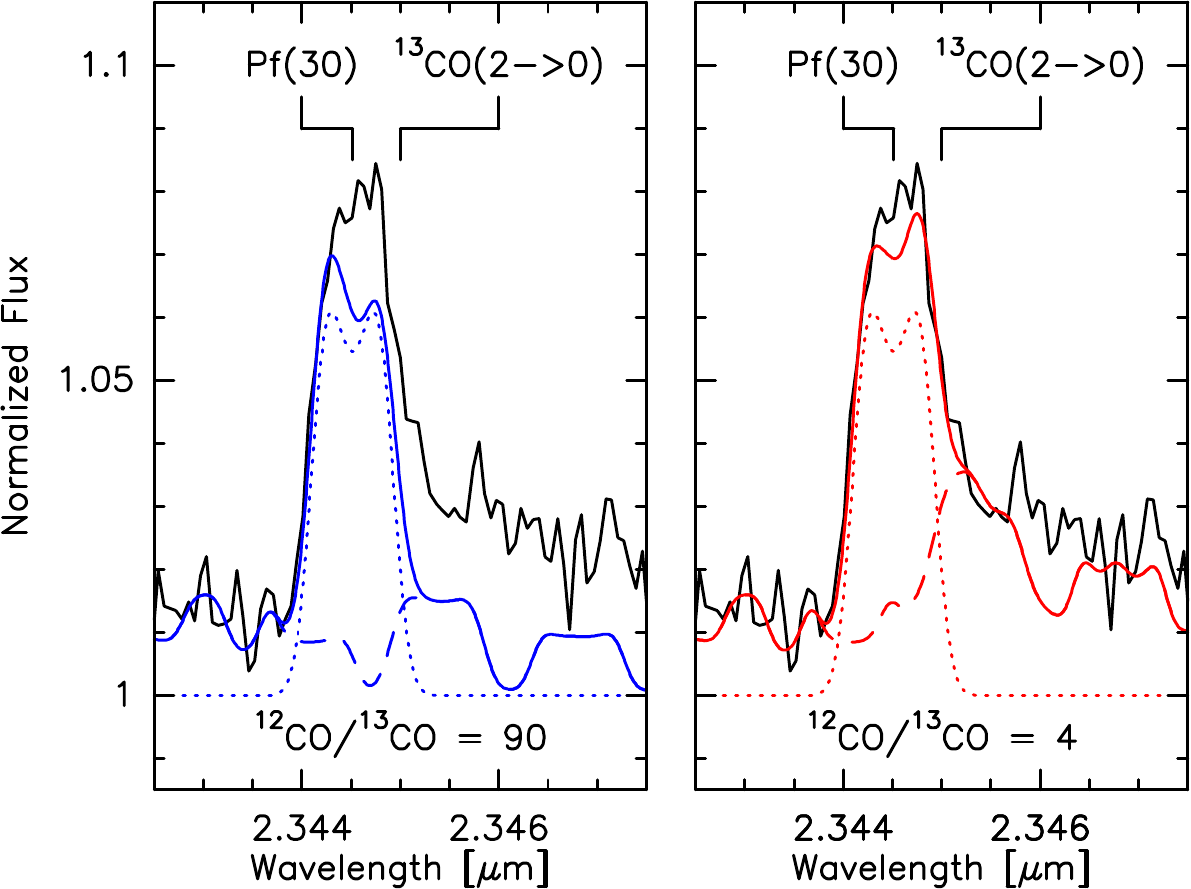}
\caption{Model fits (color) to the observations (black) zoomed to the region 
around the Pf(30) line. The dashed and dotted lines represent the pure CO and 
pure Pfund contribution to the combined fit (solid line). The fits in the left 
and right panels correspond to models with  $^{12}$CO/$^{13}$CO isotope ratios 
of 90 and 4, respectively.}
\label{fig:zoom}
\end{center}
\end{figure}

For the computation of the molecular bands we use the code developed by 
\citet{2000A&A...362..158K} for $^{12}$CO emission from a rotating disc, which 
was advanced by \citet{2009A&A...494..253K} and \citet{2013A&A...558A..17O} to 
include the bands of the isotope $^{13}$CO. The calculations are performed under 
the assumption of LTE which is a suitable approach for dense gas discs. The wide 
structure of the first band head of $^{12}$CO indicates kinematic broadening, 
although no clear hints for a blue shoulder and a red peak of the band head is 
obvious, as is typically the case for CO band emission from a rotating disc 
\citep[see, e.g.,][]{1995Ap&SS.224...25C, 2000A&A...362..158K, 
2013A&A...549A..28K, 2016A&A...593A.112K, 2015AJ....149...13M,  
2018MNRAS.480..320M}. Therefore, we follow the approach of
\citet{2000A&A...362..158K}, who proposed that the CO emission of MWC\,349A 
originates from the far side of the inner rim of the almost edge-on ($i = 
80\degr$) Keplerian disc, meaning that we compute the profiles of each 
individual CO ro-vibrational line for a symmetric ring segment with a velocity 
range spreading from $-$59\,km\,s$^{-1}$ to $+$59\,km\,s$^{-1}$. 

To fix the physical parameters of the CO emitting region, we first focus on the 
short-wavelength part of the spectrum ($\lambda <$ 2.34\,$\mu$m) which covers 
only the first two band heads of $^{12}$CO. In this wavelength region no 
contribution from $^{13}$CO arises. To reduce the number of free parameters, we 
assume a constant temperature and column density for the molecular gas. We 
compute a grid of models for the CO parameters, convolve the synthetic CO 
spectra with the resolution of the spectrograph, combine the Pfund with the CO 
emission spectra and fit these total theoretical spectra to the observations. 
From visual inspection of the fits, the best-fitting model is found for a gas 
temperature of $T_{\rm CO} =$ 1500 $\pm$ 200\,K and a $^{12}$CO column density 
of $N_{\rm CO} =$ (1.0 $\pm$ 0.5) $\times$ 10$^{21}$\,cm$^{-2}$. 

Then we include $^{13}$CO into the model. As we do not know the 
$^{12}$CO/$^{13}$CO ratio a priori, we start from the interstellar value, which 
is $\sim$90. Note, that such a small $^{13}$CO contribution remains basically 
undetectable in the total spectrum, because the intensity scales approximately 
with the column density of the gas. The fit of the combined $^{12}$CO, 
$^{13}$CO, and Pfund emission to the observed spectrum is shown as the blue line 
in Fig.\,\ref{fig:K-band-obs}.

From inspection of the fit we note that the model underestimates the observed 
intensity in the long-wavelength region ($\lambda >$ 2.344\,$\mu$m). This 
underestimation is particularly noticeable in the red peak and wing of the Pfund 
line Pf(30) at 2.3445\,$\mu$m (see left panel of Fig.\,\ref{fig:zoom}). At this 
wavelength, the only known spectral feature\footnote{We wish to 
emphasize that this feature cannot be a telluric remnant. 
Telluric lines spread over the full range of the CO bands. Improper correction 
for the telluric pollution would result in remnants all over the spectrum, which 
are not seen. In addition, the band head of $^{13}$CO is a very broad emission 
feature in contrast to the sharp telluric lines.} is the first band head of 
$^{13}$CO. The disagreement between model and observations indicates that a 
significant amount of $^{13}$CO is present in the spectrum. The best-fitting 
model is obtained for an isotope ratio of $^{12}$CO/$^{13}$CO $=$ 4 $\pm$ 1. 
This result is shown in red in the right panel of 
Fig.\,\ref{fig:zoom}. Such a high contribution of $^{13}$CO provides {\em clear 
evidence for an evolved nature of MWC\,349A} and rules out a pre-main sequence 
evolutionary state  of the star.

\section{Discussion}

We note clear differences in the physical parameters needed for the computation
of the $^{12}$CO band emission when comparing the results obtained from our 
GNIRS spectrum taken in 2013 with those derived by \citet{2000A&A...362..158K} 
from the spectrum taken in 1998 with the United Kingdom Infrared Telescope 
(UKIRT). While the $^{12}$CO column density in 1998 ($N_{\rm CO} \sim 
5 \times 10^{20}$\,cm$^{-2}$) was slightly lower but still within the errorbars 
of our value, the CO temperature ($T_{\rm CO} \simeq 3500$\,K) was 
more than twice higher in 1998 compared to the low value of $T_{\rm CO} \simeq 
1500$\,K we found now. Because of the similar resolution of the UKIRT 
($R\sim 20\,000$) and the GNIRS spectra ($R\sim 18\,000$), this discrepancy in
parameters points towards a change in excitation conditions within the disc of
MWC\,349A, which can be verified in case the two observed spectra display 
profound differences.

\begin{figure}
\begin{center}
\includegraphics[width=\hsize,angle=0]{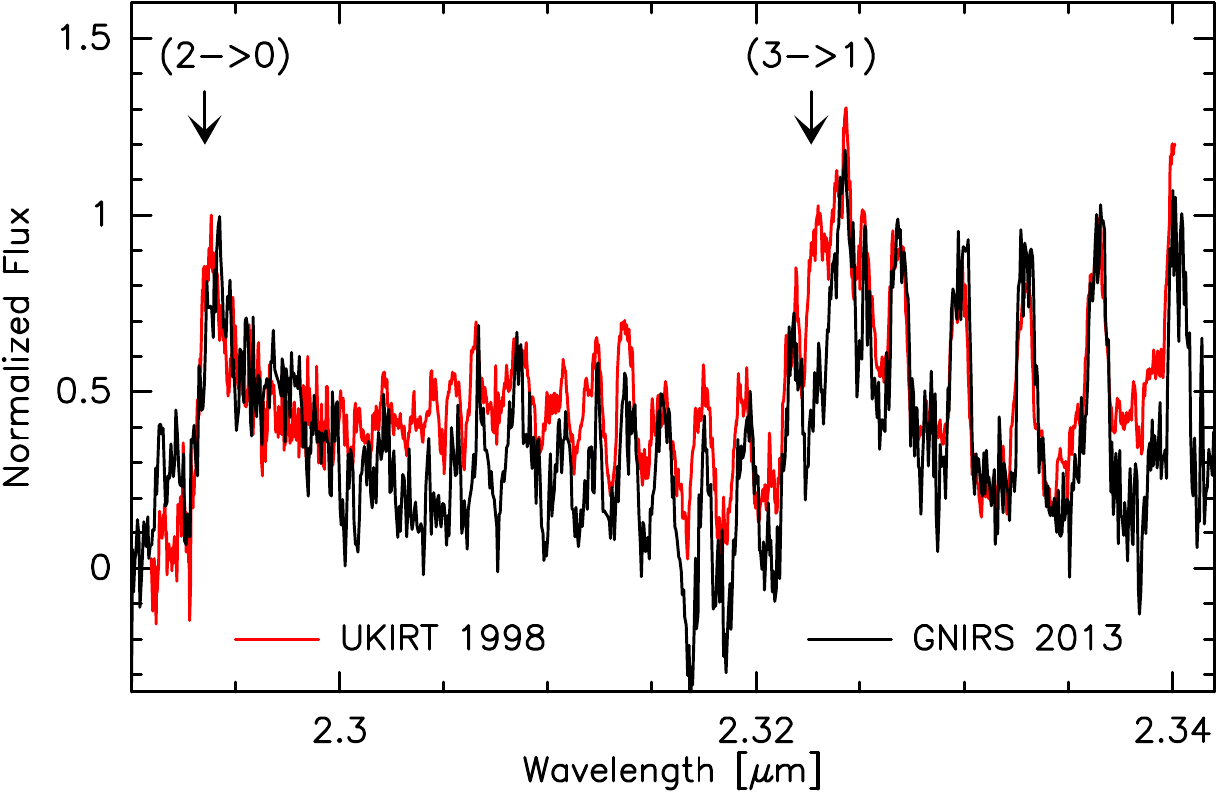}
\caption{UKIRT spectrum  taken in 1998 (red) superimposed on the  
GNIRS spectrum taken in 2013 (black). Marked are the positions of the $^{12}$CO 
band heads ($2\longrightarrow 0$) and ($3\longrightarrow 1$). For easier 
comparison, the emission is normalized to the intensity of the first band head.}
\label{fig:co_comp}
\end{center}
\end{figure}

In general, the intensity of the emission in the GNIRS spectrum is lower than in 
the UKIRT spectrum. But this trend might also be due to a variable continuum 
level and not necessarily due to changes in the CO emission. To illustrate the 
real deviations in the emission spectra, we show in Fig.\,\ref{fig:co_comp} the 
superposition of the two data sets. For easier comparison, we marked the 
positions of the band heads and normalized both spectra to the intensity of the 
first band head. The GNIRS spectrum displays less intensity between the two band 
heads, which we interpret as due to small differences in the CO column density 
along with the slightly higher velocity and the different profile shape found 
for the Pfund lines. The higher the number of the Pfund transition, the closer 
the wavelengths of the neighbouring lines, leading to strong blending of the 
lines in the upper transitions of the series and hence to a rise of the 
total emission. 

\begin{figure}
\begin{center}
\includegraphics[width=\hsize,angle=0]{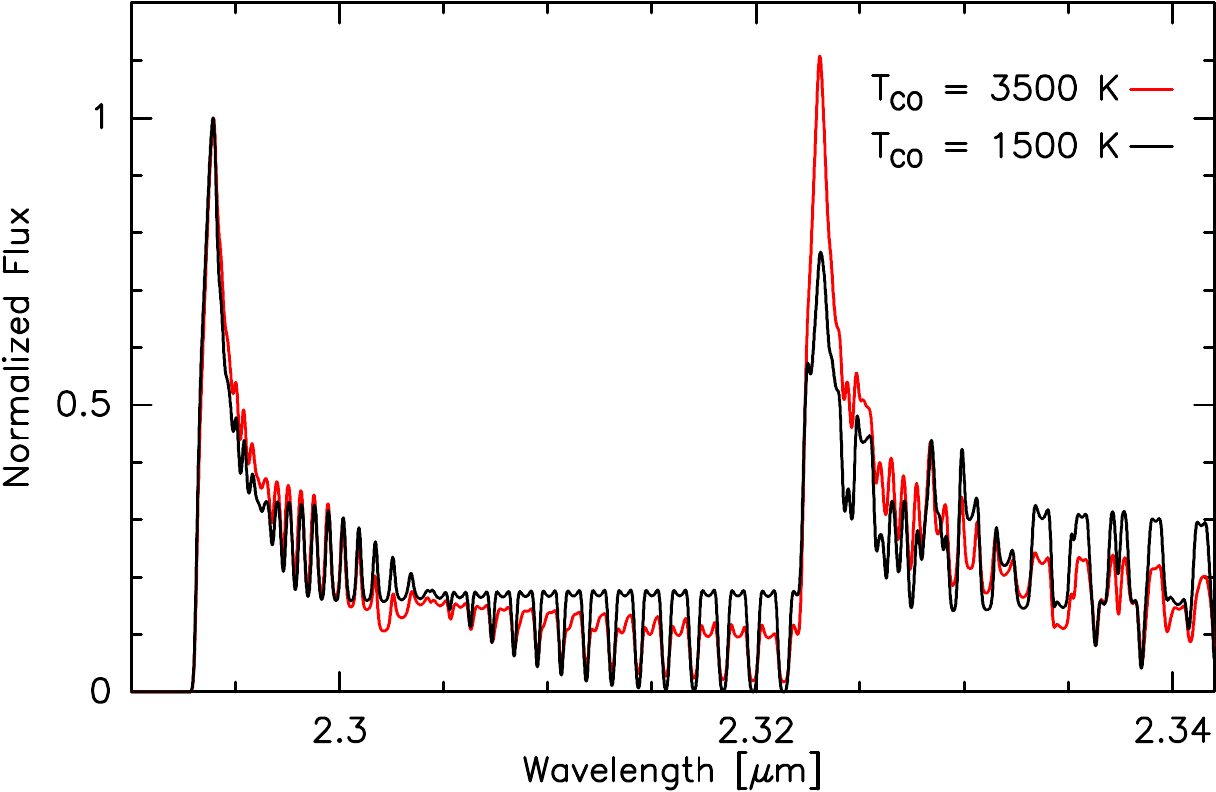}
\caption{Model spectra of pure $^{12}$CO band emission calculated for the two 
indicated temperature values. For easier comparison, the emission is normalized 
to the intensity of the first band head.}
\label{fig:co_T}
\end{center}
\end{figure}

However, the most striking change in the spectrum is the drastic decrease in the 
intensity of the ($3\longrightarrow 1$) band head. Because the excitation of the 
levels within the second band requires considerably higher energies than in 
the first band, such a vast drop in intensity is a clear indication for a 
significant decrease in gas temperature. To demonstrate the influence of the 
temperature on the intensity of the second band had, we show in 
Fig.\,\ref{fig:co_T} the synthetic spectra from CO gas with temperatures of 
3500\,K (red line) and 1500\,K (black line), whereas all other parameters were 
kept constant. The emission spectra have also been normalized to the intensities 
in the first band head. The sensitivity of the intensity of the 
($3\longrightarrow 1$) band head with temperature is obvious, reinforcing our 
conclusion that the conditions in the CO forming regions around MWC\,349A
have changed since 1998.

In this context, it is interesting to note that the first mentioning of CO band 
emission from MWC\,349A was by \citet{1987ApJ...312..297G} based on a spectrum 
taken on 1985 July 1. Despite of the low-resolution ($R \sim 650$), they clearly 
resolved the first three band heads in their data. CO band emission 
was already present in a spectrum (though not shown in the paper) taken in 1983 
by \citet{1986ApJ...311..909H}, whereas no indication of CO emission had been 
seen in the spectra that \citet{1976ApJ...205L.159T} acquired on 1975 October 13 
with an even higher resolution ($R \sim 1000$) compared to the data of 
\citet{1987ApJ...312..297G}. Another spectrum with $R \sim 1000$ had 
been taken on 1997 August 2 displaying CO band emission, also with at least 
three prominent band heads \citep{2000A&A...362..158K}, while in the new data 
from 2013 basically only the first two band heads are still visible but with 
noticeable lower intensity.

The absence of detectable CO band emission in the spectra taken in 1975 suggests 
that considerable changes in the circumstellar conditions of MWC\,349A might 
have taken place between 1975 and 1983, triggering the generation of detectable 
CO emission\footnote{Changes in the CO band spectrum can also 
arise from asymmetries or density inhomogeneities within the molecular ring. 
However, the drastic change in temperature would not fit to such a scenario, 
unless one postulates a highly eccentric ring of gas. But there is currently no 
observational evidence that would support such a hypothesis.}. These powering 
conditions seem now not to be fulfilled any more, resulting in a cooling of the
gas and an associated fading of the observable emission.  

In light of the fact that MWC\,349A is an evolved massive star, a possible 
scenario for the appearance and fading of the CO band emission might
be related to an episode with enhanced mass loss. In this regard, it is
worth mentioning that for MWC\,349A a systematic drop in visual brightness has 
been recorded in the literature during the last century. Starting with a visual 
magnitude of $13.2$\,mag \citep*{1932ApJ....76..156M}, 
\citet{1942ApJ....95..152S} reported that the star was fainter than $14$\,mag 
during their observations in 1941, while a value of $\sim 15.5$\,mag was 
obtained thirty years later by \citet{1972Natur.240..230B}. Recent photometric 
measurements reveal that the star brightened again to $V = 13$\,mag 
\citep{2017ASPC..508..389M}.

Variability in the red light with a possible period of nine years 
was reported by \citet*{2000AJ....119.3060J} based on the analysis of a sample 
of photometric plates spreading from 1967-1981.  Moreover, a change in radio 
continuum appearance from the former bipolar-shaped structure seen in the 
1980s into a square-like structure has been found by 
\citet*{2007ApJ...663.1083R} on new radio images taken in 2004. However, the 
physical cause of this morphological change is yet unclear. All these pieces 
provide clear evidence for a variable nature of MWC\,349A.

To complement our findings for MWC\,349A, we wish to mention that a sudden 
appearance of CO band emission has also been recorded for the B[e] supergiant 
LHA\,115-S\,65 in the Small Magellanic Cloud \citep{2012MNRAS.426L..56O} 
as well as for the Galactic object CI\,Cam \citep{2014MNRAS.443..947L}, in which
CO emission appeared after an outburst, and faded with the dilution of the
ejected material. 
Other B[e] supergiants were found to display fluctuations in their CO band 
emission that was ascribed to density inhomogeneities within their circumstellar 
molecular gas ring. Such a scenario has been proposed for the two B[e] 
supergiants LHA\,120-S\,73 \citep{2016A&A...593A.112K} and LHA\,120-S\,35 
\citep{2018A&A...612A.113T} in the Large Magellanic Cloud, and the Galactic 
B[e] binary object HD\,327083 \citep{2013msao.confE.160K}.

\section{Conclusions}

We have presented new high-resolution $K$-band spectra for the Galactic 
emission-line star MWC\,349A, based on which we could 
solve the long-standing issue of the star's unclear nature. The discovery of
a significant enrichment of the circumstellar material in $^{13}$CO rules out a 
pre-main sequence nature of the central star, but reinforces the classification
of MWC\,349A as evolved massive star. Proposed literature classifications are 
a B[e] supergiant \citep{1980ApJ...239..905H, 2002A&A...395..891H} or a 
luminous blue variable \citep{2012A&A...541A...7G}. Considering its near-IR 
colors, MWC\,349A appears clearly offset from luminous blue variables within the
near-IR color-color diagram, but shares its location with the B[e] supergiants
\citep{2019Galax...7...83K}. Therefore, we propose that MWC\,349A belongs to
the group of B[e] supergiants.

This finding is further supported by the variability in the CO emission 
spectrum, which seems to be inherent in B[e] supergiants. The clear decrease in 
CO intensity that we note from comparison of our new spectrum to data taken 
about 15 years ago with the same spectral resolution, is caused by a significant 
cooling of the gas from about 3500\,K down to 1500\,K. In combination with the 
fact that the brightness of MWC\,349A considerably dropped by more than 2\,mag 
between the thirties and the seventies of the last century, and that CO emission 
only started to appear in the infrared spectra of MWC\,349A after 1975, we 
further propose that the star underwent a phase of enhanced mass loss or mass 
ejection providing the environment for efficient molecule formation in order to 
generate pronounced CO band emission. With the brightening of the star back to 
its original value, it seems that the ejected material has been expanding and 
cooling, leading to the observed fading of the CO emission.

Variability in emission and in brightness seems to be an inherent property in 
this type of evolved massive stars, and a higher observing cadence is 
worthwhile to unveil the time-scales for the variabilities and their physical
origin.

\section*{Acknowledgements}

We thank the anonymous referee for valuable comments on the manuscript.
This research made use of the NASA Astrophysics Data System (ADS) and of the 
SIMBAD database, operated at CDS, Strasbourg, France. 
M.K. acknowledges financial support from GA\v{C}R (grant number 17-02337S). The 
Astronomical Institute Ond\v{r}ejov is supported by the project RVO:67985815.
M.L.A acknowledges financial support from the Programa de Incentivos (11/G160) 
of the Universidad Nacional de La Plata, Argentina. 
L.S.C. acknowledges financial support from CONICET (PIP 0177) and the Agencia
Nacional de Promoci\'{o}n Cient\'{i}ficay Tecnol\'{o}gica (PICT 2016-1971)
This project has received funding from the European Union's
Framework Programme for Research and Innovation Horizon 2020 (2014-2020)
under the Marie Sk\l{}odowska-Curie Grant Agreement No. 823734.

\bibliographystyle{mnras}
\bibliography{ms}

\bsp	
\label{lastpage}
\end{document}